\documentclass[pra,twocolumn,superscriptaddress,showpacs,10pt]{revtex4}
\usepackage{graphicx}
\usepackage{amssymb}

\begin{document}
\title{Geometric phase of an atom in a magnetic storage ring}

\author{P. Zhang}
\affiliation{School of Physics, Georgia Institute of Technology,
Atlanta, Georgia 30332, USA} \affiliation{Center for Advanced
Study, Tsinghua University, Beijing 100084, People's Republic of
China}

\author{L. You}
\affiliation{School of Physics, Georgia Institute of Technology,
Atlanta, Georgia 30332, USA} \affiliation{Center for Advanced
Study, Tsinghua University, Beijing 100084, People's Republic of
China}

\date{\today}
\begin{abstract}
A magnetically trapped atom experiences an adiabatic geometric
(Berry's) phase due to changing field direction. We investigate
theoretically such an Aharonov-Bohm-like geometric phase for atoms
adiabatically moving inside a storage ring as demonstrated in
several recent experiments.  Our result shows that this phase shift
is easily observable in a closed loop interference experiment, and
thus the shift has to be accounted for in the proposed inertial
sensing applications. The spread in phase shift due to the
atom transverse distribution is quantified through
numerical simulations.
\end{abstract}
%\pacs{05.45.Xt, 05.45.-a}
\maketitle

\section{Introduction}

A recent highlight in atom optics \cite{atomic optics} is the
realization of atomic storage rings
\cite{Chapman,Kurn1,Arnold3,Arnold2,Arnold4}. With well-designed
magnetic (B-) field configurations, cold atoms including
Bose-Einstein condensates can be confined in a ring shaped region
with a diameter of the order of a mm \cite{Kurn1}, a cm
\cite{Chapman}, or 10 cm \cite{Arnold3,Arnold2,Arnold4}. Such a
trapping geometry provides exciting prospects for delivering atoms
into high Q optical cavities \cite{Chapman} and for inertial
sensing applications \cite{Arnold2,Sagnac}.

Magnetic ring shape traps are formed by inhomogeneous magnetic
fields. Accompanied by the trapped center of mass motion, the
atomic spin adiabatically follows its local B-field direction,
giving rise to a geometric (Berry's) phase
\cite{berry,jorg,Ho-Shenoy}. This induces a phase shift in storage
ring based interference, reminiscent of the Aharonov-Bohm effect
from the magnetic gauge potential in electron interference. Even
when confined in transverse motional ground state, a distribution
of this geometric phase arises corresponding to atomic guiding
trajectories at different transverse radii, potentially causing
degradation to the interference contrast. Therefore a detailed
investigation is essential for applications of ring based
interferometry to high precision measurements.

The geometric phase of an atom in the widely used Ioffe-Pritchard
trap was first studied in Ref. \cite{Ho-Shenoy}, where the
associated gauge potential (Berry's connection)
was also discussed  for a general magnetic trap. The aim of this
paper is a detailed investigation of this geometric phase
for an atom in a magnetic storage ring. We find that it is
simply related to the cosine value of the angle between the local
B-field and the symmetry axis of the ring, a result easily
understood in terms of the solid angle sustained by the adiabatic
changing B-field along the central azimuthal closed trajectory of
a storage ring. We have performed extensive numerical studies of the
geometric phase as a function of the various components forming
the trap field and elucidated the condition when the fluctuation
of this geometric phase is small.

This paper is organized as follows: in Sec. II, we briefly review
the atomic storage ring system and discuss the associated
geometric phase. Based on the assumption of a simple classical
center of mass motion of the atom, the geometric phase and the
spatial fluctuation are calculated in Sec. III for various
types of magnetic storage rings. In Sec. IV, we
provide an full quantum treatment of this geometric phase in
a hypothetical storage ring based interference experiment.
Finally we conclude in Sec. V.

\section{Geometric phase in a storage ring}

As in recent experiments \cite{Chapman,Kurn1,Arnold3,Arnold2,Arnold4},
a magnetic storage ring can be formed by several
axially-displaced concentric current carrying coils.
Near the origin, the trap field takes the form
$\vec{B}=B_{\rho}\hat{e}_{\rho}+B_{z}\hat{e}_{z}$ \cite{Arnold1,bergman} with
\begin{eqnarray}
B_{\rho}&=&-\frac{B_{1}}{2}\rho-\frac{B_{2}}{2}\rho z,\nonumber\\
B_{z}&=&B_{0}+B_{1}z+\frac{B_{2}}{2}(z^{2}-\frac{\rho^{2}}{2}),
\label{Bfield}
\end{eqnarray}
up to polynomial terms second order in the cylindrical coordinate
$(\rho,\phi,z)$. The $B_1$ terms clearly resemble the
quadruple B-field from a pair of anti-Helmholtz coils, while the
$B_0$ and $B_2$ terms take the familiar expansions for a
Ioffe-Pritchard trap \cite{Pritchard}. It is easy to verify that
when $B_{0}B_{2}>0$, the trap field vanishes at the point
$\rho_{0}\hat{e}_{\rho}+z_{0}\hat{e}_{z}$ \cite{Arnold1} of
\begin{eqnarray}
\rho_{0}=\sqrt{4B_{0}/B_{2}-2B_{1}^{2}/B_{2}^{2}},\ \ \
z_{0}=-B_{1}/B_{2}.
\label{zero point}
\end{eqnarray}
To avoid spin flipping (Majorana transition) near this zero
B-field region, an azimuthal bias field \cite{Arnold2} can be
added, or a time averaged potential can be constructed by rapidly
oscillating the trap field $\vec{B}$ \cite{Arnold1,Kurn1}. In this
paper we mainly focus on the more complicated case of a time averaged
trap (TORT) from the recent experiment \cite{Kurn1}. When
$B_{0,1,2}$ are oscillating functions of time with a large
frequency $\omega$, the atomic motion will be governed by a time
averaged potential
$\propto(\omega/2\pi)\int_{0}^{2\pi/\omega}|\vec{B}(\vec{r},t)|dt$,
provided adiabaticity is maintained. This requires $\omega$ to be
much less than the atomic Larmor frequency in the trap field such
that its spin remains aligned. Also $\omega$ has to be much larger
than the time averaged trap frequency in order to make the
trapping effective.

For simplicity, we will suppress the $(\rho,z)$ degrees of freedom
and consider atomic motion parameterized by $\phi$ along the
azimuthal $\hat{e}_{\phi}$ direction of the storage ring. Then the
state of the atomic mass center would be described by the function
$\phi(t)$. The interaction between an atom and the presumably weak
trap field $\vec{B}[\phi,t]$ is governed by the linear Zeeman
effect
\begin{eqnarray}
H[\phi,t]=\mu_{\rm
B}g_{F}\vec{F}\cdot\vec{B}[\phi,t],
\label{Hamiltonian}
\end{eqnarray}
where $\mu_{\rm B}$ is the Bohr magneton, and $\vec{F}$ is the
atomic hyperfine spin. The {\it Lande} factor for the spin-1
manifold of a $^{87}$Rb atom is $g_{F}=-1/2$. For simplicity we
take $\hbar=1$. The trapped atomic spin is adiabatically aligned
with respect to the local B-field, i.e., the atomic spin remains
in the low field seeking internal state $|(-1)[\phi,t]\rangle_{B}$
of the instantaneous eigenstate of $H$ with eigenvalue
$\epsilon_{-1}[t]=-\mu_{\rm B}g_{F}|\vec{B}|$. It is related to
the eigenstate $|(-1)\rangle_{z}$ of $F_{z}$ (of eigenvalue $-1$)
by a simple rotation
\begin{eqnarray}
|(-1)[\phi,t]\rangle_{B}=\exp[-i\vec{F}\cdot\hat{e}_{\phi}\beta(t)]|(-1)\rangle_{z},
\label{-1F}
\end{eqnarray}
where $\beta(t)$ is the angle between $\vec{B}[\phi,t]$ and the $z$-axis.
It is independent of $\phi$ due to the cylindrical symmetry.

In this section and the following section III, we treat the atomic center of mass motion
is classically, by its one dimensional azimuthal
angle $\phi(t)$ along the circumference of the ring
(at fixed $\rho=\rho_c$ and $z=z_c$). The quantum treatment of the atomic center of mass motion
will be given in section IV.
According to the quantum adiabatic theorem \cite{adiabatic}, an
atom initially described by state
$|\Psi(0)\rangle=|(-1)[\phi(0),0]\rangle_{B}$ becomes
\begin{eqnarray}
|\Psi(t)\rangle=|(-1)[\phi(t),t]\rangle_{B}e^{-i\gamma_{d}(t)}e^{-i\gamma_{g}(t)},
\label{psi}
\end{eqnarray}
at time $t$. $\gamma_{d}=\int_{0}^{t}\epsilon_{-1}[t']dt'$
is the dynamical phase, and $\gamma_{g}(t)$ is the geometric
phase, i.e., the Berry's phase \cite{berry} that is expressed as
\begin{eqnarray}
\gamma_{g}(t)&=&-i\int_{0}^{t}\,_{B}\langle
(-1)[\phi(t'),t']|\frac
{d}{dt'}|(-1)[\phi(t'),t']\rangle_{B}dt'\nonumber\\
&=&\int_{0}^{t}\cos[\beta(t')]\Omega(t')dt',
\label{berry phase}
\end{eqnarray}
with $\Omega(t)=d\phi/dt$ the atomic angular velocity along the
storage ring. To obtain the above formula we used
$d/dt=\partial_{t}+(d\phi/dt)\partial_{\phi}$ as well as the
equalities $\,_{B}\langle (-1)[\phi,t]|
\partial_{\phi}|(-1)[\phi,t]\rangle_{B}=i\cos\beta$ and $\,_{B}\langle (-1)[\phi,t]|
\partial_{t}|(-1)[\phi,t]\rangle_{B}=0$ that come
directly from our definition of $|(-1)[\phi,t]\rangle_{B}$ in Eq.
(\ref{-1F}).

The periodic time dependence of $\vec{B}[\phi,t]$ gives rise
to the Fourier expansion
$\cos[\beta(t)]=\cos\beta_{0}+\sum_{n>0}C_{n}\cos[n\omega
t+\varphi_{n}]$ with $\cos\beta_{0}$ denoting the time averaged value
of $\cos[\beta(t)]$. On substituting into Eq. (\ref{berry phase}), we find
$\gamma_{g}=\cos\beta_{0}[\phi(t)-\phi(0)]+\sum_{n>0}C_{n}\int\cos[n\omega
t'+\varphi_{n}]\Omega(t')dt'$. The integral
$C_{n}\int\cos[n\omega t'+\varphi_{n}]\Omega(t')dt'$ takes the
maximum value $2C_{n}\Omega/(n\omega)$ for a slowly varying function $\Omega(t)$,
which becomes negligible when
$\Omega$ is much smaller than $\omega$. In this case,
we find the geometric phase $\gamma_{g}$ can be approximated as
$\gamma_{g}(t)\approx\cos\beta_{0}[\phi(t)-\phi(0)]$, which
is independent of $\Omega(t)$ and completely determined by the trajectory
of the atom in the storage ring. For a closed path, we arrive at the
geometric phase
\begin{eqnarray}
\gamma_{C}=2\pi q\cos\beta_{0},
\label{gammac}
\end{eqnarray}
where $q$ is the familiar integer winding number of the atomic
trajectory. More specifically, we find
\begin{eqnarray}
\cos\beta_{0}=\frac{\omega}{2\pi}\int_{0}^{2\pi/\omega}
\frac{B_{z}({\vec r}_{c},t)}{|\vec{B}({\vec r}_{c},t)|}dt,
\end{eqnarray}
within our approximation for various trap field configurations. In
the above, ${\vec r}_{c}=\rho_{c}\hat{e}_{\rho}+z_{c}\hat{e}_{z}$
denotes the center position of the time averaged storage ring
trap.

\section{Results}

\begin{figure}
\includegraphics[height=1.275in]{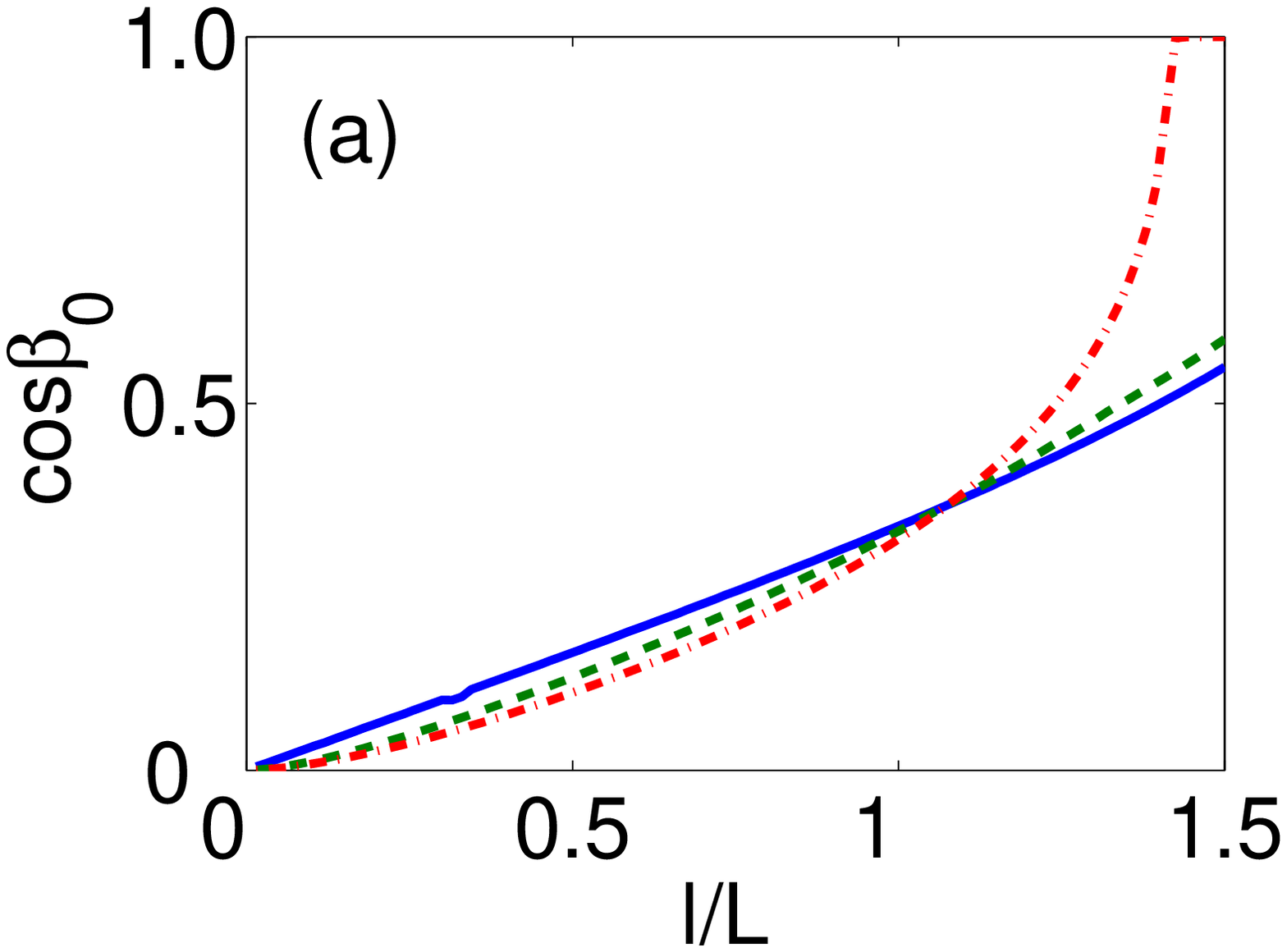}
\includegraphics[height=1.275in]{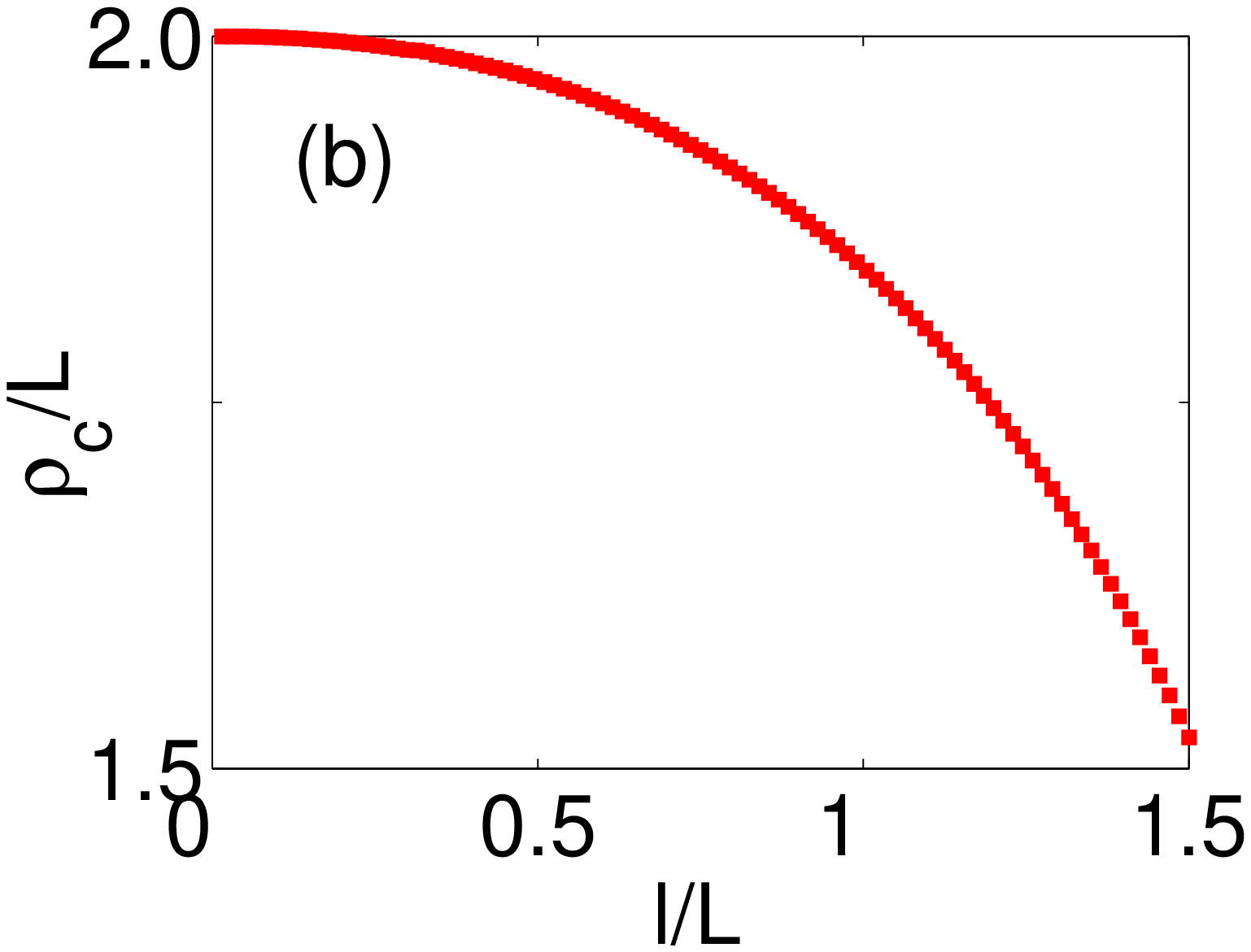}
\caption{(Color online) (a) The dependence of $\cos\beta_{0}$ on
$l/L$ for $n=0$ (solid line), $n=3L/4$ (dashed line), and $n=L$
(dash-dot line). (b) The dependence of $\rho_{c}/L$ on $l/L$ for
$n=0$. } \label{fig1}
\end{figure}

In the previous section, the general expression for the
geometric phase (\ref{gammac}) is obtained for an atom
in a storage ring. This section is devoted to a study
of the size and distribution of this
phase for various types of magnetic storage ring. For a static storage
ring trap formed by supplementing an azimuthal bias B-field from a
current carrying wire along the symmetric axis as in Ref.
\cite{Arnold2}, the Berry's phase vanishes because $\beta_0\equiv
\pi/2$. For a TORT trap, the dependence of $\cos\beta_{0}$
on the trap parameters $B_{0,1,2}$ of Eq. (\ref{Bfield}) can be
investigated. When $B_{2}$ is a constant, $\cos\beta_{0}$ is
closely related to the oscillation amplitude of $B_{1}$. For the
following simple time dependence
\begin{eqnarray}
B_{0}=B_{2}[L^{2}+n^{2}\sin(\omega t)],\ \ B_{1}=B_{2}l\cos(\omega t),
\label{oscillation B}
\end{eqnarray}
parameterized by length scales $L$, $n$, and $l$, we find that the
value of $\cos\beta_{0}$ is mainly determined by $l/L$. In Fig.
\ref{fig1}(a), the $l/L$ dependence of $\cos\beta_{0}$ is shown
for different values of $n/L$. We see that $\cos\beta_{0}$ behaves
as an increasing function of $l/L$. For $n=0$, it turns into an
approximate linear dependence $\cos\beta_{0}=0.3l/L$. In Fig.
\ref{fig1}(b), the storage ring trap center $\rho_{c}$ is shown
for $n=0$. It behaves like a quadratically decreasing function of
$l/L$.

\begin{figure}
\includegraphics[height=1.25in]{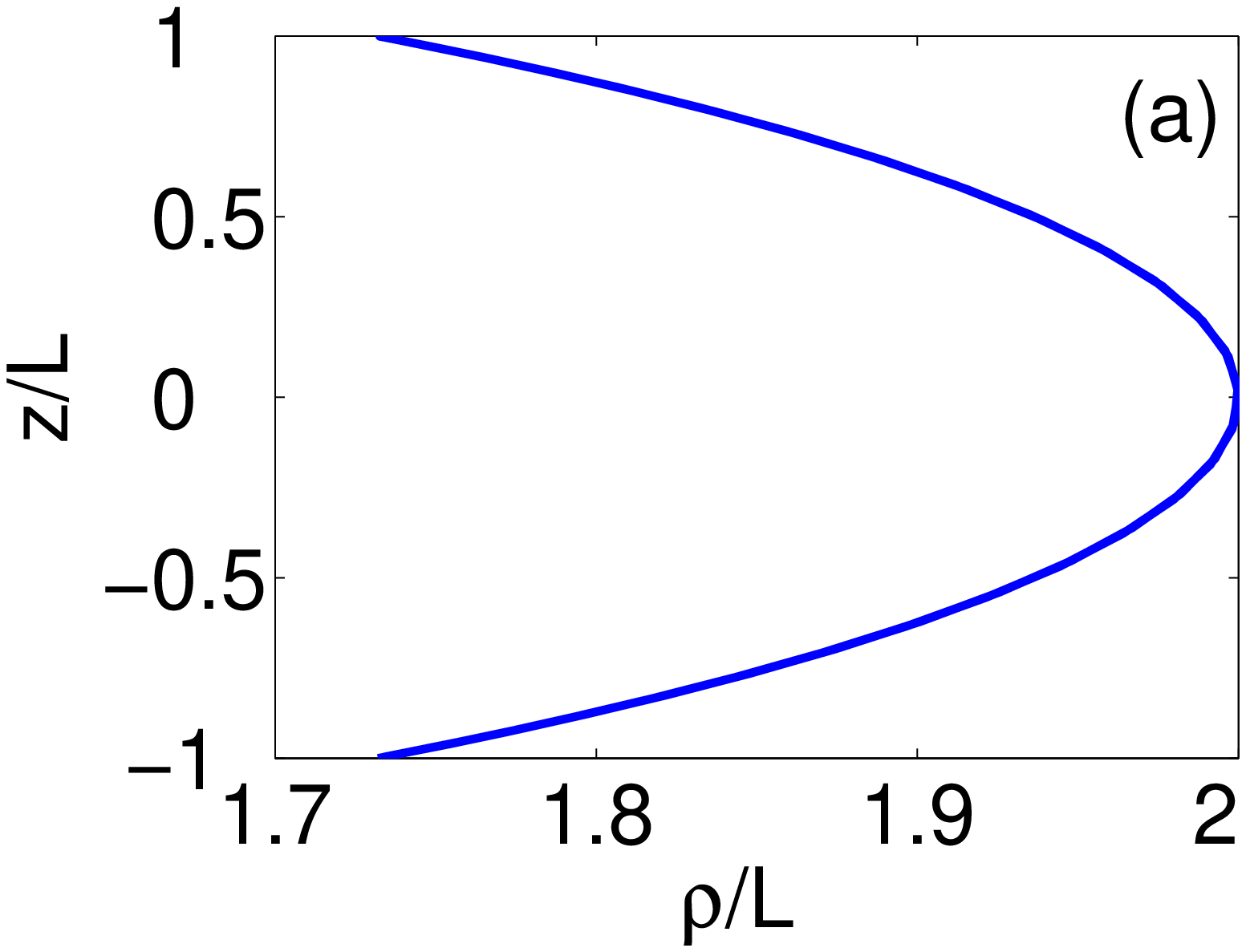}
\includegraphics[height=1.25in]{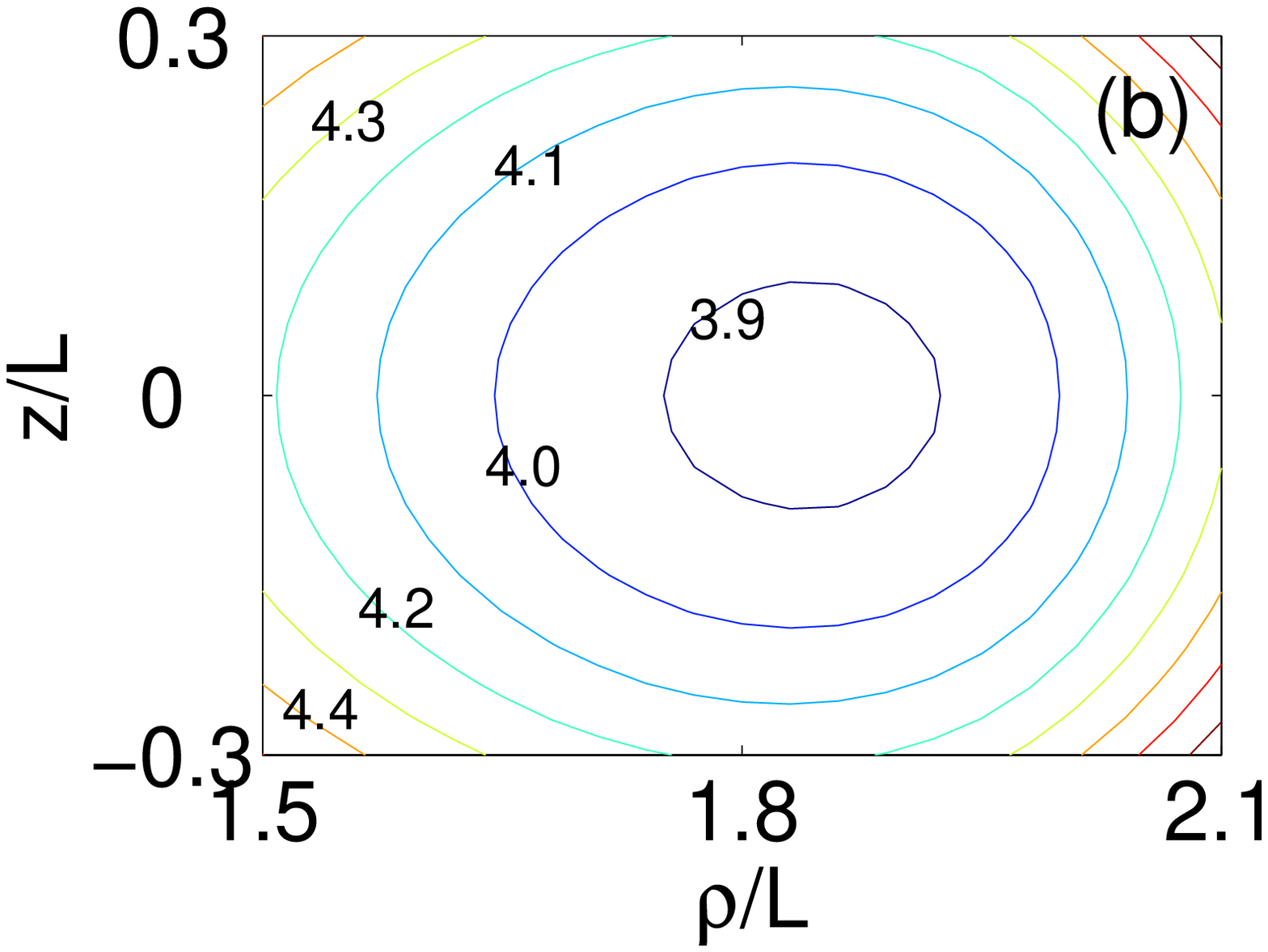}
\caption{(Color online) (a) The locus of the zero field point for
the trap's B-field. (b) The equal B-field contours of the time
averaged magnetic field. The B-field is in units of $B_{0}$. Both
figures are for $n=0$ and $l=L$.} \label{fig2}
\end{figure}

During oscillation, the zero field point of Eq. (\ref{zero point})
traces out a closed trajectory in the cross section of the storage
ring. In the proposal of Ref. \cite{Arnold1} and the experiment
\cite{Kurn1}, this closed trajectory forms a loop (a torus in 3D)
surrounding the center ($\rho_c,z_c$) along the storage ring. For
a constant $B_{2}$, the oscillation amplitude of $B_{1}$ is
proportional to the radius of this ``death torus" \cite{Kurn1} in the
$z$-direction. If the radius becomes sufficiently small, e.g., as
in Fig. 3(b) of Ref. \cite{Arnold1}, the value of $\cos\beta_{0}$
in the trap center also becomes very small. The locus of the zero
point for our example is plotted in Fig. \ref{fig2}(a). The
trajectory that forms the boundary of the loop (torus) has now
collapsed into an open curve on one side of the trap center,
although the time averaged trap potential is still well
established as illustrated in Fig. \ref{fig2}(b). We have also
calculated the Fourier coefficients $C_{n}$ of $\cos\beta(t)$. In
Fig. 3(a), the two largest coefficients $C_{2}$ and $C_{4}$ are
plotted as functions of $l/L$. We note that their maximum values
remain much smaller than unity.

\begin{figure}
\includegraphics[height=1.235in]{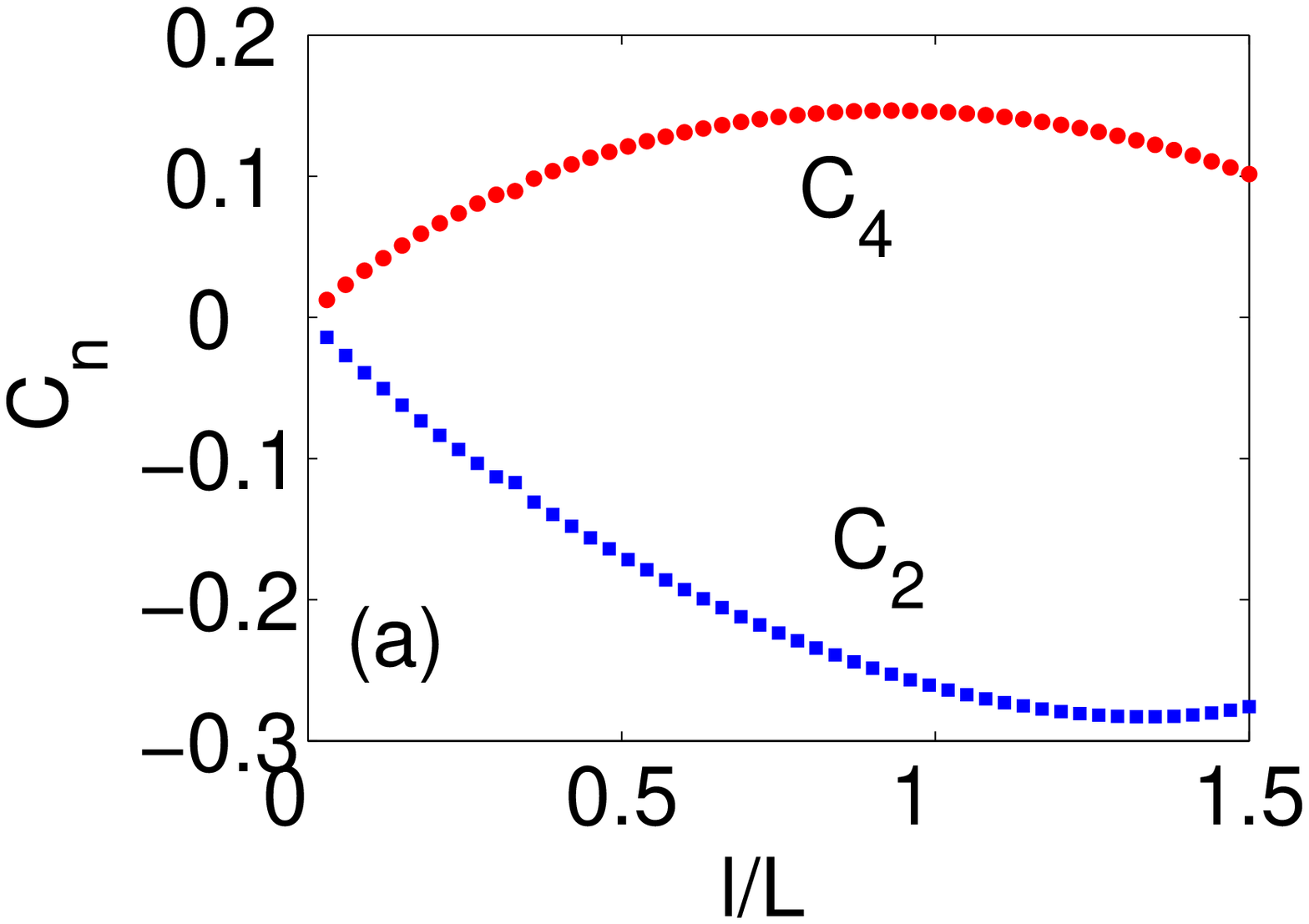}
\includegraphics[height=1.235in]{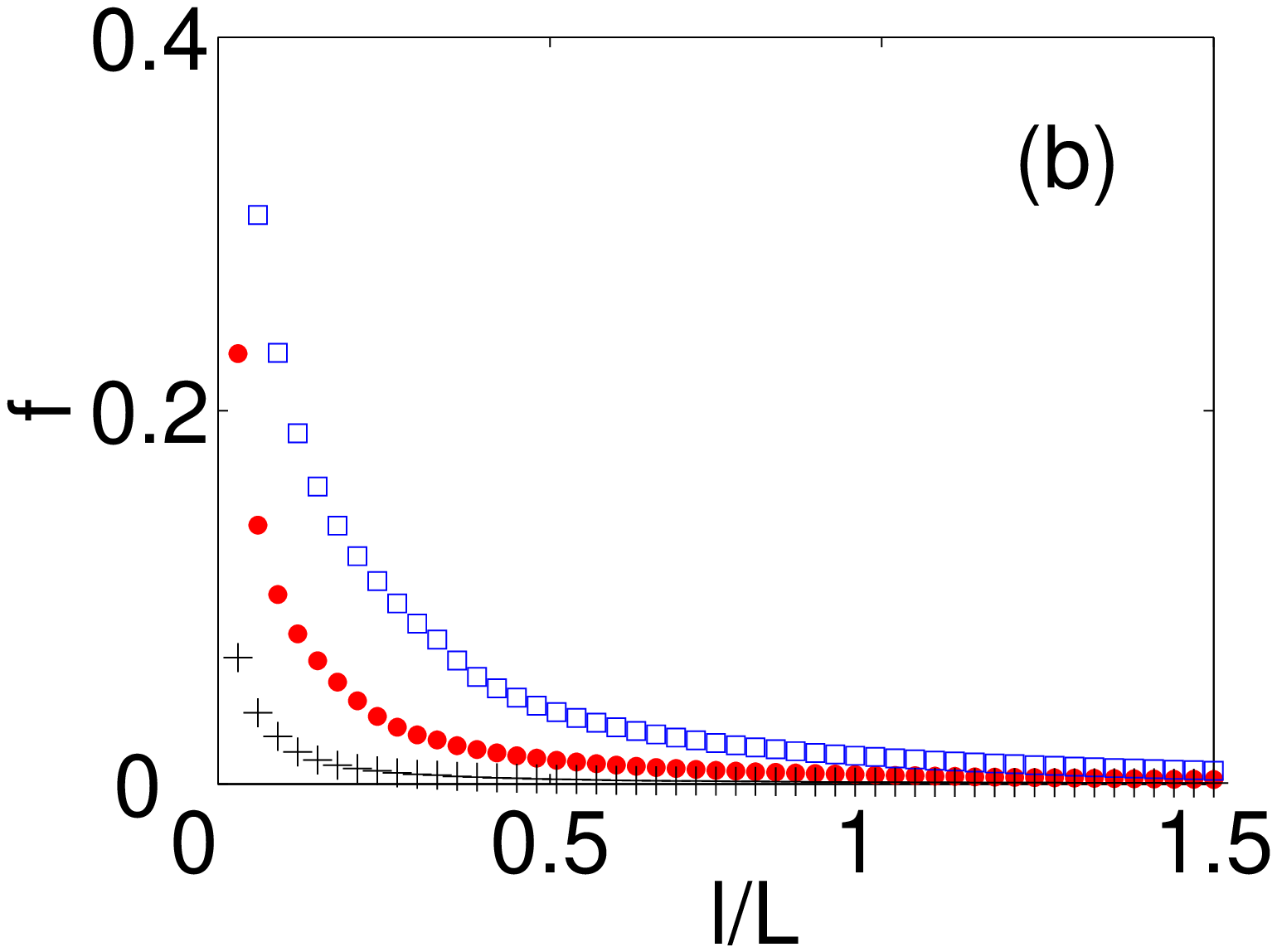}
\caption{(Color online) (a) The $l/L$ dependence of $C_{2}$
($C_{4}$) in blue (red) squares. (b) The fluctuation of
$\cos\beta_{0}$ over the region $\delta=0.015L$ (blue empty
square), $\delta=0.005L$ (red circle) and $\delta=0.001L$ (black
plus sign). } \label{fig3}
\end{figure}

In the above calculation, the classical center of mass motion of
the trapped atom along the storage ring is at a fixed ${\vec
r}_{c}$. In reality, the atomic center of mass position $(\rho,z)$
may deviate from $(\rho_{c},z_{c})$ even for a closed trajectory.
This uncertainty of atomic position in the cross sectional plane
can be due to either transverse atomic thermal motion or the
quantum nature of the transverse motional state. In the following
calculation for the fluctuations, we consider the simple case of a
transverse distribution reflected for instance by its motional
ground state. This then leads to a genuine dependence of the
geometric phase on the values of transverse coordinates $\rho$ and
$z$. This causes fluctuations of the phase shift, which can lower
interference contrast or even destroy interference fringes from
the atomic wave packet in the $\phi$ direction as we discuss in
next section. Within the semi-classical framework we adopt, the
fluctuations can be approximately simulated by a probabilistic
sampling of the values for $\rho$ and $z$ as reflected by their
probability distributions in the transverse motional ground state.
To arrive at a high contrast interference pattern, the spread of
this phase fluctuation should be kept small. We have calculated
the variations of this fluctuation with respect to different
parameters of a TORT trap. In Fig. \ref{fig3}(b), the fluctuation
defined according to $f=\sqrt{\langle
[\cos\beta_{0}(\vec{r})-\cos\beta_{0}(\vec{r}_{c})]^2\rangle}$ is
plotted as a function of $l/L$ for $\delta=0.001L$,
$\delta=0.005L$ and $\delta=0.015L$. Here $\cos\beta_{0}(\vec{r})$
is the time averaged value of $B_{z}/|B|$ at the position
$\vec{r}=\rho\hat{e}_{\rho}+z\hat{e}_{z}$. The average
$\langle\cdot\rangle$ is over a properly weighted, assumed here
for simplicity as a flat spatial distribution in the region
$R:\{\rho\in[\rho_{c}-\delta,\rho_{c}+\delta],z\in[z_{c}-\delta,z_{c}+\delta]\}$.
$f$ is found to be a decreasing function of $l/L$. Together with
Fig. \ref{fig1}(a), we conclude that when $l/L$ is large, the
value of $\cos\beta_{0}$ is also large with small fluctuations.

Finally we illustrate a practical example for
$B_{2}=7800$ Gauss$\cdot$cm$^{-2}$ and $L=l=n=0.1$ cm. The bias
field $B_{0}(t)$ is $78[1+\sin(\omega t)]$ Gauss and the first
order B-field gradient $B_{1}(t)$ becomes $780\sin(\omega
t)$ Gauss$\cdot$cm$^{-1}$. In this case the trap frequencies in the
$\rho$- and $z$- directions are $345$Hz and $672$Hz, respectively.
The time averaged Berry's phase is
$2\pi\cos\beta_{0}\approx0.2\pi$. This trap can be
realized with two pairs of Helmholtz coils (a and b) centered at
$(z=\pm A_{a,b},\rho=0)$ and a pair of anti-Helmholtz coils c
centered at $(z=\pm A_{c},\rho=0)$ with $A_{a}=0.1$cm,
$A_{b}=0.25$cm, and $A_{c}=0.5$cm, respectively. The radii
$R_{a,b,c}$ of the three pairs of coils are taken as
$R_{a}=0.3$cm, $R_{b}=0.5$cm, and $R_{c}=0.6$cm with currents in
the two Helmholtz pairs being $i_{a}=289$A and $i_{b}=-550$A, and in
the anti-Helmholtz coils are $i_{c}=335$A.

\section{The Aharonov-Bohm like phase shift from a quantum treatment}

The geometric phase as discussed above is directly measurable,
e.g., in the Sagnac interferometer \cite{Sagnac}. For this
purpose, we perform a calculation capable of a quantum treatment
of atomic motion. Trapped atoms are assumed to be in the guiding
regime so that their transverse degrees of freedom, $\rho$ and
$z$, are frozen in the ground state. Writing the atomic wave
function as $\Phi(\phi,t)|(-1)(\phi,t)\rangle_{B}$ with
$\Phi(\phi,t)$ for its spatial motion, we find that the
Schrodinger equation $i(d/dt)\Psi=H_{\rm ad}\Psi$ is governed by a
time averaged adiabatic Hamiltonian \cite{sunandother}
\begin{eqnarray}
H_{\rm
ad}=\frac{1}{2m}{\left[p_{\phi}-{A(\phi)}/{\rho_c}\right]^{2}}+V(\phi).
\label{quantumHamiltonian}
\end{eqnarray}
$V(\phi)$ denotes the time averaged trap potential that more
generally can contain additional optical or the gravitational
potentials. $A(\phi)=-i(\omega/2\pi)\int_{0}^{2\pi/\omega}\
_{B}\langle(-1)(\phi,t)|\partial_{\phi}|(-1)(\phi,t)\rangle_{B}dt$
is the time averaged gauge potential. For the simple case
considered previously, $A$ equals $\cos\beta_{0}$ and is
independent of $\phi$.
In this section, without loss of generality, we assume
$A(\phi)$ to be a slowly varying function of $\phi$. The operator
$p_{\phi}=-i\rho_c^{-1}\partial/\partial_{\phi}$ denotes the
atomic canonical momentum along the $\hat{e}_{\phi}$ direction. In
the presence of the gauge potential $A(\phi)$, the atomic kinetic
angular velocity $v_{\phi}$ becomes
$v_{\phi}=p_{\phi}/(m\rho_{c})-A(\phi)/(m\rho_{c}^{2})$.

We note that the
Hamiltonian in Eq. (\ref{quantumHamiltonian}) only applies
when the gauge potential $A$ do not
depend strongly on $\rho$ and $z$, as otherwise their
dependence in the term
$A(\phi,\rho,z)p_{\phi}$ would lead to inseparable correlations
between the dependencies on $\phi$, $\rho$, and $z$ of the
atomic wave function. If this does occur, the reduced
atomic quantum state for the $\phi$ direction becomes
a mixed state rather than a pure one, resulting in reduced
coherence property. This constitutes a quantum explanation
for the reduced contrast in the interference pattern due to
fluctuations of the geometric phase.

Obviously this Hamiltonian (\ref{quantumHamiltonian}) resembles
the motion of an electron around a current carrying solenoid,
where an Aharonov-Bohm (A-B) phase \cite{AB effect} is induced by
the corresponding vector potential, with a value proportional to
its integral along a closed path surrounding the solenoid. In our
case, we expect essentially the same result with an
Aharonov-Bohm-like phase, or the Berry's phase $\gamma_{C}$ as
defined in Eq. (\ref{gammac}), given by the line integral of the
induced gauge potential $\int A(\phi)d\phi$. In the A-B effect of
an electron, both the electron source and the detector (screen)
are far away from the solenoid, where the vector gauge potential
vanishes. The electron achieves a steady asymptotic distribution
long before and after ($t\to\pm\infty$) the encounter with the
solenoid. Therefore its motion can be treated via stationary
scattering theory \cite{AB effect,ABBerry} or based on a path integral
approach with an incident plane wave \cite{cohen}. For our case of
trapped atoms in a storage ring, the gauge potential $A(\phi)$
never vanishes at any instant during the evolution. Thus our model
must be treated dynamically as described by a time
dependent Schrodinger equation.

We first consider the motion of a single wave packet assuming that
$A(\phi)$ varies slowly and is approximated as a constant in the
region, where the wave function is non-negligible. The expectation
value of the atomic position $\phi$, the canonical momentum
$p_{\phi}$, and the velocity $v_{\phi}$ of the wave packet at time $t$
will be denoted as $\bar{\phi}_{t}$, $\bar{p}_{t}$, and
$\bar{v}_{t}$. At $t=0$ the wave packet $\varphi(\phi,0)$ can be
written as
\begin{eqnarray}
\varphi(\phi,0)=\psi(\phi,0)e^{iA(\bar{\phi}_{0})(\phi-\bar{\phi}_{0})}.
\label{initial wave packet}
\end{eqnarray}
Both $\bar{\phi}_{0}$ and $\bar{v}_{0}$ are determined completely through
$\psi(\phi,0)$,
\begin{eqnarray}
\bar{\phi}_{0}&=&\int \phi|\psi(\phi,0)|^{2}d\phi,\nonumber\\
\bar{v}_{0}&=&-i{1\over {m\rho_c}}\int
\psi^{*}(\phi,0)\partial_{\phi}\psi(\phi,0)d\phi,
\end{eqnarray}
independent of the gauge function $A$.

The Hamiltonian $H_{\rm ad}$ (\ref{quantumHamiltonian}) can be
obtained from the one without gauge potential $A(\phi)$ via a
transformation, i.e., $H_{\rm ad}=UH_{0}U^{-1}$ with
\begin{eqnarray}
H_{0}=\frac{p_{\phi}^{2}}{2m}+V(\phi),\
U=\exp[i\int_{\bar{\phi}_{0}}^{\phi}A(\phi')d\phi'],
\end{eqnarray}
provided the wave packet is limited to and moving in a finite
region over the circumference. Therefore, $\varphi(\phi,t)$ is
written as
\begin{eqnarray}
\varphi(\phi,t)&\approx&\psi(\phi,t)e^{iA(\bar{\phi}_{t})
(\phi-\bar{\phi}_{t})}e^{i\int_{\bar{\phi}_{0}}^{\bar{\phi}_{t}}A(\phi')d\phi'},
\label{wavepacket}
\end{eqnarray}
with $\psi(\phi,t)=e^{-iH_{0}t}\psi(\phi,0)$, where we have
used the relation
$e^{-iHt}=Ue^{-iH_{0}t}U^{-1}$ and replaced $A(\phi)$ with
$A(\bar{\phi}_{0})$ [$A(\bar{\phi}_{t})$] in the region where
$\psi(\phi,0)$ [$\psi(\phi,t)$] is non-negligible.
The center position (velocity)
$\bar{\phi}_{t}$ ($\bar{v}_{t}$) of each wave packet
and the variation of the profile
$\psi$ are also determined by $H_{0}$ independent of $A(\phi)$.

\begin{figure}
\includegraphics[width=2.35in,height=1.75in]{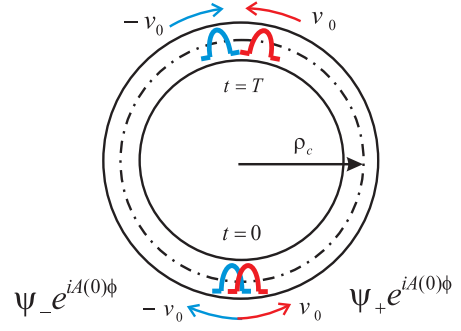}
\caption{(Color online) A proposed setup of interferes
in a storage ring. An initial superposition of
two wave packets $\psi_{\pm}(\phi)e^{iA(0)\phi}$
with counter propagating velocities $\pm v_{0}$
interferences at time $T$ when the packets
overlap near the other end $\bar{\phi}^{(+)}_{T}$.}
\label{fig4}
\end{figure}

To detect the geometric phase, we consider an interference
set up as shown in Fig. \ref{fig4}
with the initial atomic state being a superposition
of two counter propagating wave packets $\psi_{\pm}e^{iA(0)\phi}$
centered at $\phi=0$
\begin{eqnarray}
\Phi(\phi,0)=\frac{1}{\sqrt{2}}\left[\psi_{+}(\phi,0)
e^{iA(0)\phi}+\psi_{-}(\phi,0)e^{iA(0)\phi}\right].
\label{initial}
\end{eqnarray}
The two wave packets overlap again at time $T$ when
$\bar{\phi}^{(+)}_{T}=\bar{\phi}^{(-)}_{T}+2\pi$. Making use of
Eq. (\ref{wavepacket}), we find
%\begin{widetext}
\begin{eqnarray}
\Phi(\phi,T)=\frac{1}{\sqrt{2}}[\psi_{+}(\phi,T)
e^{iA(\bar{\phi}^{(+)}_{T})(\phi-\bar{\phi}^{(+)}_{T})}
e^{i\int_{0}^{\bar{\phi}^{(+)}_{T}}A(\phi')d\phi'}&&\nonumber\\
+\psi_{-}(\phi-2\pi,T)
e^{iA(\bar{\phi}^{(+)}_{T})(\phi-\bar{\phi}^{(+)}_{T})}
e^{i\int_{0}^{\bar{\phi}^{(+)}_{T}-2\pi}A(\phi')d\phi'}],&& \hskip 12pt
\end{eqnarray}
%\end{widetext}
with $\psi_{\pm}(\phi,T)=e^{-iH_{0}T}\psi_{\pm}(\phi,0)$. A
substitution of $\phi\rightarrow\phi-2\pi$ to the counter
propagating wave packet defines the complete wave function over
the same azimuthal region $(0,2\pi)$. An interference pattern
$\propto\cos[\eta_{+}(\phi)-\eta_{-}(\phi)+\gamma_{C}]$ then
reveals the presence of the phase shift $\gamma_{C}$, where
$\eta_{+}$ ($\eta_{-}$) is defined as the phase of
$\psi_{+}(\phi,T)$ [$\psi_{-}(\phi-2\pi,T)$]. Since the evolution
of $\psi_{\pm}$ is governed by $H_{0}$, the density distribution
$n_{\pm}(\phi)$ for each component and the phase difference
$\eta_{+}(\phi)-\eta_{-}(\phi)$ are independent of $A(\phi)$. The
induced gauge potential causes a shift, simply the geometric phase
shift $\gamma_{C}$, to the interference pattern.

In many situations, the center of mass motion for the
two interfering paths of an atom can be described classically.
In this case the
interference pattern can be predicted with more detail.
We show in the appendix that
Eq. (\ref{wavepacket}), which describes the
motion of a single wave packet, can be generalized to
the following
\begin{eqnarray}
\varphi(\phi,t)\approx e^{iS_{\rm
cl}[\bar{\phi}_{t},t;\bar{\phi}_{0},0]}e^{i[\bar{p}_{t}-\bar{p}_{0}][\phi-\bar{\phi}_{t}]}
e^{-i[\bar{\phi}_{t}-\bar{\phi}_{0}]p_{\phi}}\varphi(\phi,0),\nonumber\\
\label{semiclassical}
\end{eqnarray}
with $\bar{\phi}_{t}$ and $\bar{p}_{t}$ denoting the
expectation values for
the position and its canonical momentum that satisfy the Hamiltonian
equation governed by $H_{\rm ad}$, i.e., the center
velocity $\bar{v}_{t}$ of the wave packet satisfies
\begin{eqnarray}
\bar{v}_{t}=\bar{p}_{t}-A(\bar{\phi}_{t})=d\bar{\phi}_{t}/dt.
\end{eqnarray}
As we have pointed out before, the classical trajectory $\bar{\phi}_{t}$
and the speed
$\bar{v}_{t}$ are independent of the gauge potential
$A(\phi)$ because the dynamics is one dimensional.
The global phase $S_{\rm cl}$ is defined as the action
function in classical mechanics:
\begin{eqnarray}
S_{\rm
cl}(\bar{\phi}_{t},t;\bar{\phi}_{0},0)=\int_{0}^{t}L[\bar{\phi}_{t'},\bar{v}_{t'}]dt',
\end{eqnarray}
where
$L[\bar{\phi}_{t},\bar{v}_{t}]=\bar{v}_{t}^{2}/2+A(\bar{\phi}_{t})\bar{v}_{t}-V(\bar{\phi}_{t})$
is the Lagrangian function in classical mechanics. With this
semiclassical description for the wave packet inside the storage
ring, we find that the relative phase
$[\eta_{+}(\phi)-\eta_{-}(\phi)+\gamma_{C}]$ can be approximated
more transparently as
$[m\rho_{c}(\bar{v}^{(+)}_{T}-\bar{v}^{(-)}_{T})\phi+\xi+\gamma_{C}]$
with a gauge-independent constant
\begin{eqnarray}
\xi&=&\eta_{0+}(0)-\eta_{0-}(0)+\int_{0}^{\bar{\phi}^{(+)}_{T}}[(\rho_{c}\bar{v}_{t}^{(+)})^{2}/2-V(\bar{\phi}^{(-)}_{t})]dt\nonumber\\
&&-\int_{0}^{\bar{\phi}^{(-)}_{T}}[(\rho_{c}\bar{v}_{t}^{(-)})^{2}/2-V(\bar{\phi}^{(-)}_{t})]dt.
\end{eqnarray}

The discussions outlined in this section also allow for easy
calculations of more general Aharonov-Bohm like phase shifts
due to different gauge effects in storage ring based
matter wave interference experiments. For example,
the matter wave Sagnac effect, which can be considered as an
analog of the Aharonov-Bohm effect \cite{sagnac-2},
is easily deduced to correspond to a
phase shift $2\pi m\Omega\rho_{c}^{2}\cos\theta$ in a
storage ring set up, with $\Omega$ being the angular velocity
for rotation and $\theta$ being the
angle between the axis of rotation and the $z$ direction.
The deduction simply involves replacing the adiabatic gauge
protectional $A(\phi,t)$ with $m\Omega\rho_{c}^{2}\cos\theta$,
which is proportional to the $\phi$-component of the "gauge field"
$m\vec{\Omega}\times\vec{r}$ for the Sagnac
effect \cite{sagnac-2}.

\section{Conclusion}

In this paper, we have presented the theoretical analysis of the
geometric phase for a neutral atom inside a magnetic storage ring.
For the interesting example of a time averaged potential (TORT),
we have shown that the geometric phase is proportional to the
averaged cosine value of the angle between the local B-field and
the $z$ axis. When the oscillation amplitude of the quadruple
component $B_{1}$ is large, a significant and sharply peaked
geometric phase is realized on completing a single
pass along the storage ring. Of course the net effect is
cumulative with respect to the number of turns an atom makes. We
hope our result will shine new light on the proposed inertial
sensing experiments based on trapped atoms in storage rings.

\begin{acknowledgments}
%put your acknowledgments here.
We thank Mr. Bo Sun, Dr. Duanlu Zhou, Prof. Lee Chang, and Prof. C.
P. Sun for helpful discussions. This work is supported by NASA,
NSF, and CNSF.
\end{acknowledgments}

\appendix
\section{The One Dimensional Semi-Classical Motion of a Single Wave Packet}

During the course of this work, we obtained some interesting
results on the semi-classical one-dimensional motion
for an atomic wave packet. These are summarized here because
to our knowledge, they do no seem to have been stated or
known anywhere.

Essentially the following material provides
a proof of Eq. (\ref{semiclassical}) which
describes the one dimensional semi-classical motion of a single
wave packet $\varphi(\phi,t)$. Our calculation is based on the
following two assumptions:
\begin{enumerate}
\item During the whole evolution, the wave packet is assumed to be
sufficiently narrow that
the potential $V(\phi)$ can be approximated by a linear function
in $\phi$ and the gauge potential $A_{\phi}$ assumed a
constant in the region where the wave packet is non-negligible;

\item The evolution time is so short that the effect of
wave packet diffusion can be omitted.
\end{enumerate}

With the first assumption, it is easy to prove \cite{cohen quantum}
that the center position $\bar{\phi}_{t}$, the center
canonical momentum $\bar{\phi}_{t}$, and the center velocity
$\bar{v}_{t}$ satisfy the classical Hamiltonian equations
\begin{eqnarray}
{d\bar{\phi}_{t} \over dt}&=&\bar{v}_{t}=\bar{p}_{t}-A(\bar{\phi}_{t}),\nonumber\\
{d\bar{p}_{t}\over dt}&=&-\frac{\partial }{\partial \bar{p}_{t}}H_{\rm
ad}(\bar{\phi}_{t},\bar{p}_{t}).\label{Hamiltonian equation}
\end{eqnarray}

As we have shown in Eq. (\ref{initial wave packet}), at $t=0$,
the wave packet can be written as
$\varphi(\phi,0)=\psi(\phi,0)e^{iA(\bar{\phi}_{0})(\phi-\bar{\phi}_{0})}$.
Furthermore, we can express the profile $\psi(\phi,0)$ as
\begin{eqnarray}
 \psi(\phi,0) =\int dp
f(p,0)e^{-ip(\phi-\bar{\phi}_{0})}e^{-i\bar{v}_{0}(\phi-\bar{\phi}_{0})},
\label{psi0}
\end{eqnarray}
where the function $f(p,0)$ describes the momentum distribution at
$t=0$.

At time $t$, the atomic wave function $\varphi(\phi,t)$ is
expressed as in Eq. (\ref{wavepacket}), determined by $
\psi(\phi,t)=e^{-iH_{0}t} \psi(\phi,0)$.
To calculate the wave function $\psi(\phi,t)$,
we first replace the trap potential $V(\phi)$ with a
time dependent linear potential $V_{0}(t)+V_{1}(t)\phi$ with
\begin{eqnarray}
V_{0}(t)&=&V(\bar{\phi}_{t})-\frac{dV(\phi)}{d\phi}\Big|_{\phi=\bar{\phi}_{t}}\bar{\phi}_{t},\nonumber\\
V_{1}(t)&=&\frac{dV(\phi)}{d\phi}\Big|_{\phi=\bar{\phi}_{t}}.\nonumber
\end{eqnarray}
The
Hamiltonian equation (\ref{Hamiltonian equation}) satisfied by the
center position and velocity can now be casted in the form of
a Newtonian equation
\begin{eqnarray}
d^{2}\bar{\phi}_{t}/dt^{2}&=&-V_{1}(t),\nonumber \\
d\bar{\phi}_{t}/dt&=&\bar{v}_{t},
\end{eqnarray}
which can be solved as
\begin{eqnarray}
\bar{\phi}_{t}&=&\bar{\phi}_{0}+\bar{v}_{0}t-\int_{0}^{t}dt_{1}\int_{0}^{t_{1}}dt_{2}V_{1}(t_{2}),\nonumber\\
\bar{v}_{t}&=&\bar{v}_{0}-\int_{0}^{t}dt_{1}V_{1}(t_{1}).\label{classical
orbit}
\end{eqnarray}

According to the first assumption, the wave function
$\psi(\phi,t)=e^{-iH_{0}}\psi(\phi,0)$ satisfies the Schoredinger
equation
\begin{eqnarray}
i\partial_{t}\psi(\phi,t)=H_{\rm eff}(t)\psi(\phi,t),
\label{effective schoredinger equation}
\end{eqnarray}
 approximately, with the
effective Hamiltonian
\begin{eqnarray}
H_{\rm
eff}(t)=\frac{p_{\phi}^{2}}{2}+V_{0}(t)+V_{1}(t)\phi.
\end{eqnarray}

Because the effective Hamiltonian $H_{\rm eff}$ is a quadratic
function, the Schr\"odinger Eq. (\ref{effective schoredinger
equation}) can be solved with the Wei-Norman algebra method
\cite{wei-norman}. After direct calculation, we find
\begin{eqnarray}
\psi(\phi,t)={\cal U}(t)\psi(\phi,0),
\label{effective unitary}
\end{eqnarray}
with the evolution operator given by
\begin{eqnarray}
{\cal
U}(t)=e^{-ia(t)p_{\phi}^{2}}e^{-ib(t)p_{\phi}}e^{-ic(t)\phi}e^{-id(t)},
\label{unitary}
\end{eqnarray}
parameterized by
\begin{eqnarray}
\begin{array}{lll}
a(t)&=&t/2,\\
b(t)&=&\int_{0}^{t}dt_{1}t_{1}V_{1}(t_{1}),\\
c(t)&=&\int_{0}^{t}dt_{1}V_{1}(t_{1}), \\
d(t)&=&\int_{0}^{t}dt_{1}V_{0}(t_{1})+\int_{0}^{t}dt_{1}\int_{0}^{t_{1}}dt_{2}V_{1}(t_{1})t_{2}V_{1}(t_{2}).
\end{array}
\label{abcd}
\end{eqnarray}

In the end, we find
\begin{eqnarray}
\psi(\phi,t) =e^{iS_{0}(t)}\int dp
f(p,t)e^{-ip(\phi-\bar{\phi}_{t})}e^{-i\bar{v}_{t}(\phi-\bar{\phi}_{t})},\ \ \
\label{psit}
\end{eqnarray}
with $f(p,t)=f(p,0)e^{-ip^{2}t/2}$ and the global phase factor
\begin{eqnarray}
S_{0}(t)=\int_{0}^{t}L_{0}[\bar{\phi}_{t'},\bar{v}_{t'}]dt',
\end{eqnarray}
with $L_{0}[\bar{\phi}_{t},\bar{v}_{t}]=\bar{v}_{t}^{2}/2-V_{0}(t)-V_{1}(t)\bar{\phi}_{t}$.
In the above derivation, we have used Eq. (\ref{classical orbit}) and the
relationship
$\int_{0}^{t}dt_{1}(t-t_{1})V_{1}(t_{1})=\int_{0}^{t}dt_{1}\int_{0}^{t_{1}}dt_{2}(t-t_{1})V_{1}(t_{2})$.

If $t$ is so small that the momentum uncertainty
$\Delta p$ of the wave function $\psi(\phi,0)$ remains small,
we can omit the wave packet diffusion and assume
$f(p,t)\approx f(p,0)$, which is our second assumption above.
With this approximation and substituting Eq. (\ref{psit}) into Eq.
(\ref{wavepacket}), we obtain Eq. (\ref{semiclassical}).
Of course, this same result can also be deduced
with the method in Sec. 3 of Ref. \cite{sun}.

\end{document}